\documentclass{article}
\usepackage[T1]{fontenc} 
\usepackage[utf8]{inputenc} 
\usepackage{amsmath,cite,url}
\usepackage{graphicx}
\usepackage[lbd]{ismir}
\usepackage{amssymb}

\usepackage{color}

\usepackage[firstpage]{draftwatermark}
\definecolor{lightgray}{rgb}{0.9,0.9,0.9}
\definecolor{darkgray}{rgb}{0.4,0.4,0.4}
\SetWatermarkFontSize{12pt}
\SetWatermarkScale{1.1}
\SetWatermarkAngle{90}
\SetWatermarkHorCenter{202mm}
\SetWatermarkVerCenter{170mm}
\SetWatermarkColor{darkgray}
\SetWatermarkText{Late-Breaking / Demo Session Extended Abstract, ISMIR 2025 Conference}

\title{Exploring State-Space-Model based Language Model in Music Generation}

\multauthor
{Wei-Jaw Lee \hspace{1cm} Fang-Chih Hsieh \hspace{1cm} Xuanjun Chen} { \bfseries{Fang-Duo Tsai \hspace{1cm} Yi-Hsuan Yang \hspace{1cm}}\\
 Graduate Institute of Communication Engineering, National Taiwan University, Taiwan\\
{\tt\small weijaw2000@gmail.com}
}

\def\authorname{Wei-Jaw Lee, Fang-Chih Hsieh, Xuanjun Chen, Fang-Duo Tsai, and Yi-Hsuan Yang}

\begin{document}

\maketitle
\begin{abstract}
The recent surge in State Space Models (SSMs)~\cite{gu2023mamba, dao2024mamba2}, particularly the emergence of Mamba, has established them as strong alternatives or complementary modules to Transformers across diverse domains. In this work, we aim to explore the potential of Mamba-based architectures for text-to-music generation. We adopt discrete tokens of Residual Vector Quantization (RVQ) as the modeling representation and empirically find that a single-layer codebook can capture semantic information in music. Motivated by this observation, we focus on modeling a single-codebook representation and adapt SiMBA, originally designed as a Mamba-based encoder, to function as a decoder for sequence modeling. We compare its performance against a standard Transformer-based decoder. Our results suggest that, under limited-resource settings, SiMBA achieves much faster convergence and generates outputs closer to the ground truth. This demonstrates the promise of SSMs for efficient and expressive text-to-music generation. We put audio examples on Github.\footnote{https://lonian6.github.io/web-exploring-ssm/}
\end{abstract}
\section{Introduction}\label{sec:introduction}

Recent text-to-music (TTM) generation models have exhibited impressive capabilities in generating high-quality audio from text descriptions. The existing state-of-the-art (SOTA) models predominantly adopt diffusion \cite{liu2024audioldm2,melechovsky2023mustango,evans2025stable,li2024jen1,lanzendorfer2025coarse,tsai25musecontrollite, wu2024musiccontrolnet} or Transformer-based \cite{agostinelli2023musiclm,copet2023simple, yang2023uniaudio} backbones. The former approach tends to use a variational autoencoder (VAE) to compress an audio waveform into a continuous latent space, whereas the latter employs an RVQ-based audio codec model 
to encode a waveform into discrete tokens.

To explore the potential of Mamba in music generation, we adopt a similar approach to the transformer-based model proposed in MusicGen~\cite{copet2023simple}, given that both Mamba and Transformer operate in an auto-regressive manner. Specifically, we use Descript Audio Codec (DAC)~\cite{kumar2023dac} as the audio codec and train language models (LMs) based on both SiMBA and Transformer architectures under the same training configurations. We empirically find that modeling only a single-layer codebook leads to suboptimal audio fidelity, but doing so is sufficient to preserve most of the semantic content in music. Therefore, as a preliminary study, we restrict our modeling to the first-layer codebook of DAC to reduce the model complexity. 

The main contributions of this work are as follows:
1) Our work represents one of the first attempts to employ SiMBA as a decoder architecture for text-to-music generation, aiming to explore the potential of SSM-based architectures for audio generation.
2) We demonstrate that SiMBA achieves faster convergence and better generation quality compared to a Transformer-based baseline under fair training conditions.


\section{Methodology}
\label{sec_method}

\subsection{Motivation for Single-Codebook Modeling}
Some Transformer-based models, such as MusicGen~\cite{copet2023simple}, utilize RVQ-based audio codec models to encode audio as a sequence of discrete tokens.
To ensure high audio fidelity, each short audio segment is typically encoded recursively using $K$ quantization codebooks $\{Q_1, \dots, Q_K\}$. Consequently, a waveform $\mathbf{w}$ is represented as a token matrix $x \in \mathbb{R}^{K \times L}$, where $K$ is the number of codebooks and $L$ is the frame length.

\begin{figure}
  \centering
  \includegraphics[alt={Language model architectures},width=\linewidth,trim=0 {0.1\linewidth} 0 0]{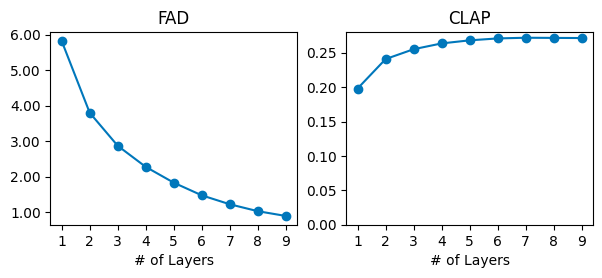}
  \caption{\textbf{The coarse-to-fine property of audio tokens}: the audio quality (measured in FAD) and text-audio alignment (measured in CLAP) of the ``reconstructed'' audio encoded and decoded by the DAC~\cite{kumar2023dac} using the first $\kappa$ quantization layers (from 1 to 9) over the MusicCaps dataset.
  }
  \label{fig:score_1}
\end{figure}

To investigate the contribution of different RVQ layers to audio reconstruction, we conduct a preliminary analysis using DAC tokens~\cite{kumar2023dac}, which consist of $K = 9$ quantization layers and operate at a frame rate of approximately 86 frames per second.
We use DAC to encode each audio sample in the MusicCaps dataset~\cite{agostinelli2023musiclm} into a sequence $x \in \mathbb{R}^{K \times L}$, where $K = 9$, and reconstruct it using the first $\kappa$ layers of tokens, where $\kappa \in {1, ..., 9}$.
We evaluate each reconstructed audio with different layers using Fréchet Audio Distance (FAD)\cite{kilgour2018fr} (lower is better) for audio quality and CLAP\cite{elizalde2023clap} (higher is better) for text-audio alignment.
As shown in Figure~\ref{fig:score_1}, we find that audio reconstructed using only the coarsest layer of tokens retains a significant amount of semantic information. However, achieving higher audio quality requires incorporating more layers. 

\subsection{SiMBA}

\begin{figure}
  \centering
  \includegraphics[alt={Language model  architectures},width=0.8\linewidth]{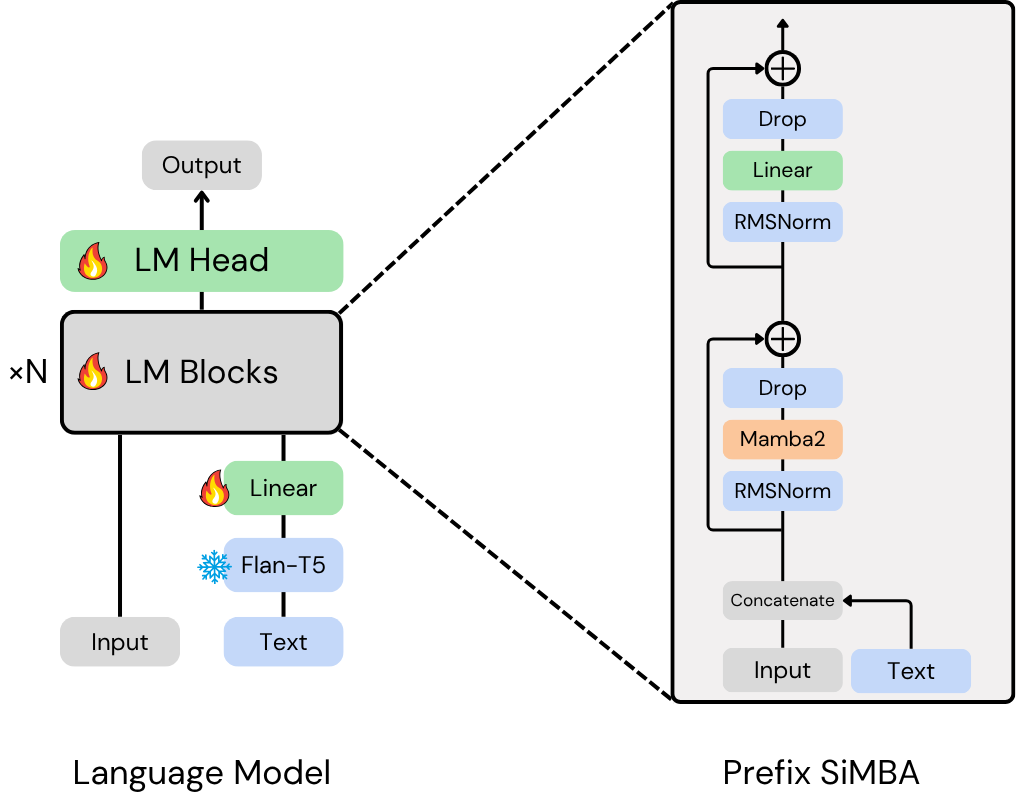}
  \caption{\textbf{LM architectures.} Left: Overall architecture illustrating the input RVQ audio tokens and the use of Flan-T5~\cite{chung2024t5} as a pre-trained text encoder. Right: The SiMBA~\cite{patro2024simba} architecture conditioned on the prefix text.
  }
  \label{fig:LM_model}
\end{figure}

Figure~\ref{fig:LM_model} depicts the overall LM and the SiMBA layer architecture. It has two inputs, the preceding audio tokens for the first coarse layer, and the sequence of input text prompt $c$, which is encoded into embedding vectors using a pretrained  Flan-T5 Base model \cite{chung2024t5}. 
We have a cascade of $N$ ``LM blocks'' modeling the temporal dependency of the audio tokens and the text-audio dependency. 
All the LM components except for the Flan-T5 are set to be trainable.
While the original SiMBA uses the more sophisticated Einstein FFT (EinFFT)~\cite{patro2024simba} 
for channel mixing, we opt for a simpler structure, adopting the basic linear layers here. Aside from the SiMBA, we also train a vanilla Transformer as a comparable baseline with cross attention to condition the generation on the input text, named Cross Transformer.

\section{Experimental Setup}
\label{section:Experimental Results}
Our training data is from Jamendo~\cite{bogdanov2019mtg}.
We re-sample all the audio into 44.1kHz and convert them into mono audio, splitting the tracks into non-overlapping 30s clips with vocals removed by HTDemucs~\cite{rouard2023hybrid}. 
LP-MusicCaps~\cite{doh2023lp} generates per clip three 10s captions, rephrased by LLaMA-3~\cite{grattafiori2024llama} into a single description, excluding clips with residual vocal cues. Moreover, to prevent data leakage, we exclude tracks overlapping with the Song Describer Dataset (SDD)~\cite{manco2023sdd}. We further discard clips with more than 15s of silence and randomly select a clip per track for training to ensure musical richness. 

In our work, we implement the DAC 44.1kHz model with about 86 frame rate. We use Flan-T5 Base model as text encoder with 768 embedding dimension. The training processes are on a single RTX 3090 GPU. The hidden dimension of LM blocks is 1,024, a Mamba-2 block state dimension of 512 with a dropout rate of 0.3. While training, the batch size is set to 4 with gradient accumulation every 32 steps. We use the AdamW optimizer~\cite{loshchilov2017adamW} with learning rate 1e--4 and weight decay 2e--2, betas (0.9, 0.999), a cosine annealing scheduler and a 100-step warm-up.

\section{Results and Conclusion}

\begin{figure}
  \centering
  \includegraphics[alt={LM performance along the training steps},width=0.8\linewidth,trim=0 {0.1\linewidth} 0 0]{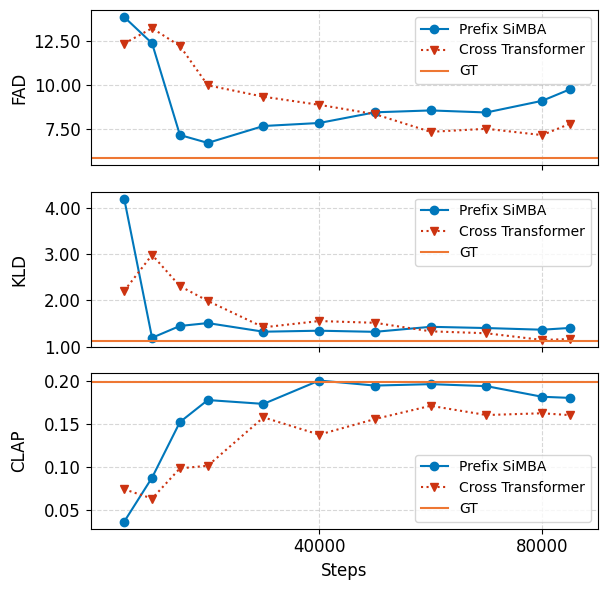}
  \caption{Objective evaluation result of Prefix SiMBA and Cross Transformer for 10s generation as training unfolds.}
  \label{fig:score_training_steps}
\end{figure}

We conduct objective evaluation using the following metrics: {Fréchet Audio Distance (FAD)}\cite{kilgour2018fr}, {Kullback-Leibler divergence (KLD)}, and {CLAP}\cite{elizalde2023clap}, which respectively measure audio quality, semantic similarity, and text-audio alignment. Both FAD and KLD are lower the better, while CLAP is higher the better. Figure~\ref{fig:score_training_steps} illustrates the 10-second music generation performance across three metrics over the course of 85k training steps.
We observe that Prefix SiMBA generally outperforms the Cross Transformer baseline during the early training stage (i.e., before 40k steps). As training progresses, Prefix SiMBA continues to achieve higher CLAP and comparable KLD. However, it exhibits a slightly higher FAD compared to the baseline. 

In sum, our preliminary results demonstrate that Prefix SiMBA, as a decoder architecture for audio generation, exhibits faster convergence and superior semantic alignment compared to a Transformer-based baseline during early training. 
Even after extended training, SiMBA maintains competitive performance in terms of semantic similarity and text-audio alignment. 
These findings highlight the potential of state space model (SSM)-based architectures for efficient and effective music generation.

\bibliography{ISMIR2025_template}

\begin{thebibliography}{10}
\providecommand{\url}[1]{#1}
\csname url@samestyle\endcsname
\providecommand{\newblock}{\relax}
\providecommand{\bibinfo}[2]{#2}
\providecommand{\BIBentrySTDinterwordspacing}{\spaceskip=0pt\relax}
\providecommand{\BIBentryALTinterwordstretchfactor}{4}
\providecommand{\BIBentryALTinterwordspacing}{\spaceskip=\fontdimen2\font plus
\BIBentryALTinterwordstretchfactor\fontdimen3\font minus \fontdimen4\font\relax}
\providecommand{\BIBforeignlanguage}[2]{{%
\expandafter\ifx\csname l@#1\endcsname\relax
\typeout{** WARNING: IEEEtran.bst: No hyphenation pattern has been}%
\typeout{** loaded for the language `#1'. Using the pattern for}%
\typeout{** the default language instead.}%
\else
\language=\csname l@#1\endcsname
\fi
#2}}
\providecommand{\BIBdecl}{\relax}
\BIBdecl

\bibitem{loshchilov2017adamW}
I.~Loshchilov and F.~Hutter, ``Decoupled weight decay regularization,'' in \emph{Int. Conf. Learning Representations}, 2019.

\bibitem{bogdanov2019mtg}
D.~Bogdanov, M.~Won, P.~Tovstogan, A.~Porter, and X.~Serra, ``{The MTG-Jamendo Dataset for Automatic Music Tagging},'' in \emph{Proc. Machine Learning for Music Discovery Workshop, Int. Conf. Machine Learning}, 2019.

\bibitem{rouard2023hybrid}
S.~Rouard, F.~Massa, and A.~D{\'e}fossez, ``Hybrid {Transformers} for music source separation,'' in \emph{Proc. IEEE Int. Conf. Acoustics, Speech and Signal Processing}, 2023.

\bibitem{doh2023lp}
S.~Doh, K.~Choi, J.~Lee, and J.~Nam, ``{LP-MusicCaps}: {LLM}-based pseudo music captioning,'' in \emph{Proc. International Society for Music Information Retrieval Conference}, 2023.

\bibitem{grattafiori2024llama}
A.~Grattafiori, A.~Dubey, A.~Jauhri, A.~Pandey, A.~Kadian, A.~Al-Dahle, A.~Letman, A.~Mathur, A.~Schelten, A.~Vaughan \emph{et~al.}, ``The {LLaMA 3} herd of models,'' \emph{arXiv preprint arXiv:2407.21783}, 2024.

\bibitem{manco2023sdd}
I.~Manco, B.~Weck, S.~Doh, M.~Won, Y.~Zhang, D.~Bogdanov, Y.~Wu, K.~Chen, P.~Tovstogan, E.~Benetos, E.~Quinton, G.~Fazekas, and J.~Nam, ``{The Song Describer Dataset}: a corpus of audio captions for music-and-language evaluation,'' in \emph{Machine Learning for Audio Workshop at NeurIPS 2023}, 2023.

\bibitem{chung2024t5}
H.~W. Chung, L.~Hou, S.~Longpre, B.~Zoph, Y.~Tay, W.~Fedus, Y.~Li, X.~Wang, M.~Dehghani, S.~Brahma \emph{et~al.}, ``Scaling instruction-finetuned language models,'' \emph{Journal of Machine Learning Research}, vol.~25, no.~70, pp. 1--53, 2024.

\bibitem{gu2023mamba}
A.~Gu and T.~Dao, ``Mamba: Linear-time sequence modeling with selective state spaces,'' \emph{arXiv preprint arXiv:2312.00752}, 2023.

\bibitem{dao2024mamba2}
T.~Dao and A.~Gu, ``Transformers are {SSM}s: Generalized models and efficient algorithms through structured state space duality,'' in \emph{Proc. Int. Conf. Machine Learning}, 2024.

\bibitem{patro2024simba}
B.~N. Patro and V.~S. Agneeswaran, ``{SiMBA}: Simplified mamba-based architecture for vision and multivariate time series,'' \emph{arXiv preprint arXiv:2403.15360}, 2024.

\bibitem{kumar2023dac}
R.~Kumar, P.~Seetharaman, A.~Luebs, I.~Kumar, and K.~Kumar, ``High-fidelity audio compression with improved {RVQGAN},'' \emph{Advances in Neural Information Processing Systems}, vol.~36, pp. 27\,980--27\,993, 2023.

\bibitem{lanzendorfer2025coarse}
L.~A. Lanzend{\"o}rfer, T.~Lu, N.~Perraudin, D.~Herremans, and R.~Wattenhofer, ``Coarse-to-fine text-to-music latent diffusion,'' in \emph{Proc. IEEE Int. Conf. Acoustics, Speech and Signal Processing}, 2025.

\bibitem{agostinelli2023musiclm}
A.~Agostinelli, T.~I. Denk, Z.~Borsos, J.~Engel, M.~Verzetti, A.~Caillon, Q.~Huang, A.~Jansen, A.~Roberts, M.~Tagliasacchi \emph{et~al.}, ``{MusicLM: Generating music from text},'' \emph{arXiv preprint arXiv:2301.11325}, 2023.

\bibitem{copet2023simple}
J.~Copet, F.~Kreuk, I.~Gat, T.~Remez, D.~Kant, G.~Synnaeve, Y.~Adi, and A.~D{\'e}fossez, ``Simple and controllable music generation,'' \emph{Advances in Neural Information Processing Systems}, vol.~36, pp. 47\,704--47\,720, 2023.

\bibitem{evans2025stable}
Z.~Evans, J.~D. Parker, C.~Carr, Z.~Zukowski, J.~Taylor, and J.~Pons, ``Stable audio open,'' in \emph{Proc. IEEE Int. Conf. Acoustics, Speech and Signal Processing}, 2025.

\bibitem{liu2024audioldm2}
H.~Liu, Y.~Yuan, X.~Liu, X.~Mei, Q.~Kong, Q.~Tian, Y.~Wang, W.~Wang, Y.~Wang, and M.~D. Plumbley, ``{AudioLDM 2: Learning Holistic Audio Generation With Self-Supervised Pretraining},'' \emph{IEEE/ACM Transactions on Audio, Speech, and Language Processing}, vol.~32, pp. 2871--2883, 2024.

\bibitem{melechovsky2023mustango}
J.~Melechovsky, Z.~Guo, D.~Ghosal, N.~Majumder, D.~Herremans, and S.~Poria, ``Mustango: Toward controllable text-to-music generation,'' in \emph{Proc. Conf. North American Chapter of the Association for Computational Linguistics: Human Language Technologies}, 2024.

\bibitem{li2024jen1}
P.~P. Li, B.~Chen, Y.~Yao, Y.~Wang, A.~Wang, and A.~Wang, ``{JEN-1: Text-guided universal music generation with omnidirectional diffusion models},'' in \emph{Proc. IEEE Conference on Artificial Intelligence}, 2024, pp. 762--769.

\bibitem{yang2023uniaudio}
D.~Yang, J.~Tian, X.~Tan, R.~Huang, S.~Liu, X.~Chang, J.~Shi, S.~Zhao, J.~Bian, X.~Wu \emph{et~al.}, ``Uniaudio: An audio foundation model toward universal audio generation,'' \emph{arXiv preprint arXiv:2310.00704}, 2023.

\bibitem{wu2024musiccontrolnet}
S.-L. Wu, C.~Donahue, S.~Watanabe, and N.~J. Bryan, ``{Music ControlNet}: Multiple time-varying controls for music generation,'' \emph{IEEE/ACM Transactions on Audio, Speech, and Language Processing}, vol.~32, pp. 2692--2703, 2024.

\bibitem{kilgour2018fr}
K.~Kilgour, M.~Zuluaga, D.~Roblek, and M.~Sharifi, ``{Fr\'{e}chet Audio Distance}: A metric for evaluating music enhancement algorithms,'' \emph{arXiv preprint arXiv:1812.08466}, 2018.

\bibitem{elizalde2023clap}
B.~Elizalde, S.~Deshmukh, M.~Al~Ismail, and H.~Wang, ``{CLAP}: Learning audio concepts from natural language supervision,'' in \emph{Proc. IEEE Int. Conf. Acoustics, Speech and Signal Processing}, 2023.

\bibitem{tsai25musecontrollite}
F.-D. Tsai, S.-L. Wu, W.~Lee, S.-P. Yang, B.-R. Chen, H.-C. Cheng, and Y.-H. Yang, ``{MuseControlLite}: Multifunctional music generation with lightweight conditioners,'' in \emph{Proc. Int. Conf. Machine Learning}, 2025.

\end{thebibliography}

%
%
%
%
%

\end{document}